\documentstyle[11pt,newpasp,twoside]{article} 

\def\edcomment#1{\iffalse\marginpar{\raggedright\sl#1\/}\else\relax\fi}
\reversemarginpar

\def\kms{km ${\rm s}^{-1}$}

\def\ch2{$\chi^2$}

\def\kms {\hbox{${\rm km\ s}^{-1}$}}

\def \HI {H{\sc \,i}}

\def\lapp{\ifmmode\stackrel{<}{_{\sim}}\else$\stackrel{<}{_{\sim}}$\fi}
\def\gapp{\ifmmode\stackrel{>}{_{\sim}}\else$\stackrel{>}{_{\sim}}$\fi}

\begin{document}
\title{Deep Searches for High Redshift Molecular Absorption}
\author{S. J. Curran and J. K. Webb}
\affil{School of Physics, University of New South Wales, NSW 2052, Australia}
\author{M. T. Murphy}
\affil{Institute of Astronomy, Madingley Road, Cambridge CB3 0HA, UK}
\author{N. Kuno}
\affil{Nobeyama Radio Observatory, Nagano 384-1305, Japan} 
  
\begin{abstract}
Millimetre-band scans of the frequency space towards optically dim
quasars is potentially a highly efficient method for detecting new
high redshift molecular absorption systems. Here we describe scans
towards 7 quasars over wide bandwidths (up to 23 GHz) with sensitivity
limits sufficient to detect the 4 redshifted absorbers already
known. With wider frequency bands, highly efficient searches of
large numbers of possibly obscured objects will yield many new
molecular absorbers.

\end{abstract}

\section{Introduction}

Webb et al. (these proceedings) discussed constraints on possible
variations in fundamental constants offered by quasar absorption lines. Optical studies
(Webb et al. 1999; Murphy et al. 2003) find a
statistically significant variation of the fine-structure constant,
$\Delta\alpha/\alpha \approx (-0.54 \pm 0.12)\times 10^{-5}$, over the
redshift range $0.2 < z_{\rm abs} < 3.7$.  Comparison between \HI-21cm and
molecular rotational (millimetre) absorption lines can yield an order of
magnitude better precision (per absorption system) than these purely
optical constraints (Drinkwater et al. 1998; Carilli et al. 2000; Murphy
et al. 2001): a statistical sample of \HI-21cm/mm comparisons will provide
an important cross-check on varying-$\alpha$. Currently, however, only 4
such redshifted millimetre absorption systems are known (Wiklind, these
proceedings). To increase this number we have employed the following
search strategies:

\begin{enumerate}
  \item Deep integrations of damped Lyman-alpha absorbers (DLAs),
    the highest column density ($N_{\rm HI}\gapp10^{20}$ cm$^{-2}$)
    quasar absorbers known. Since we observe at a known redshift
    and therefore frequency, optical depth limits better than $\tau\lapp0.1$
    are often obtained. The DLA results are discussed in detail by Curran et al. (2004).
    \item Scanning the frequency space towards visually dim millimetre
      bright quasars in search of a possible absorber responsible for
      the visual obscuration. Here we summarise our results as obtained
      with the Swedish-ESO Sub-millimetre Telescope (SEST) and 
      Nobeyama Radio Observatory's 45-m telescope (NRO).
\end{enumerate}
\section{Results}
From an extensive literature search we selected four millimetre-loud
quasars yet to be optically identified (Table 1, top).
\begin{table}
\begin{center}
\caption{The SEST (top) and NRO (bottom) search results. $V$ is
the visual magnitude with the Galactic extinction, $A_B$,
given. $z_{\rm em}$ is the quasar redshift, $S$ the approximate
flux density in Jy at the observed frequency band, $\nu$, and $\tau$ is the
typical $3\sigma$ optical depth limit (quoted for a resolution of 1 \kms, see Curran et al. 2002).}
\begin{tabular}{l c c c c c   c  c }			
\noalign{\smallskip}
\tableline
\noalign{\smallskip}
Quasar & $V$ & $A_B$ & $z_{\rm em}$ & Ref & $S$ & $\nu$ [GHz] & $\tau$ \\
\noalign{\smallskip}
\tableline
\noalign{\smallskip}
0500+019 & 21.2 & 0.289 & 0.58457 & 2 & 0.5 & 78.30--80.90 & 2\\
...&...& ...& ...&...&0.5 & 85.50--86.50 & 0.8\\
...&...& ...& ...&...&0.5 & 112.10--113.10 & 0.8\\
...&...& ...& ...&...&0.4 & 130.00--141.40 & 1 \\
0648--165 & -- & 2.456 & -- & -- & 1.6 & 78.30--80.90 & 0.2 \\
...&...& ...& ...&...&1.5 & 83.90--90.50 & 0.2\\
...&...& ...& ...&...&0.7 & 138.00--141.40 & 0.5 \\
...&...& ...& ...&...&0.6 & 142.80--149.40 & 0.7 \\
0727--115 & 22.5 & 1.271 & -- & -- & 2.9 & 78.30--80.90 & 0.2 \\
...&...& ...& ...&...& 2.7 & 83.90--90.50 & 0.1 \\
...&...& ...& ...&...&1.5 & 138.00--141.40 & 0.2 \\
...&...& ...& ...&...&1.2 & 142.80--148.60 & 0.4 \\
1213--172 & 21.4 & 0.253 & -- & -- &  1.0& 78.30--80.90 & 0.4 \\
...&...& ...& ...&... & 0.9 & 83.90--90.50 & 0.4 \\
...&...& ...& ...&... & 0.5 &138.00--141.40 & 0.8 \\
...&...& ...& ...&... & 0.4& 142.80--151.00 & 1\\
\noalign{\smallskip}
\tableline
\noalign{\smallskip}
0742+103 & $\sim24$ &0.111 & -- & --& 0.6 & 46.90--47.50  & --\\
...&...& ...& ...&...&0.4&77.25--87.50 & 0.9\\
...&...& ...& ...&...&0.4&88.45--89.35 & 0.4\\
1600+335 & 23.2& 0.137 &1.1 & 3 &0.7 &46.90--47.50 & 2\\
...&...& ...&...&...&0.5& 77.25--87.50 &0.5\\
...&...& ...&...&...&0.5& 88.45--89.35 &0.4\\
1655+077 & 20.1 &0.66& 0.621 & 1 &1.5 &46.90--47.50 &1 \\
...&...& ...& ...&...&1.2&77.25--87.50 &0.7\\ 
...&...& ...& ...&...&1.2&88.45--89.35 &0.3\\ 
\noalign{\smallskip}
\tableline
\noalign{\smallskip}
\end{tabular}
{References: (1) Wilkes (1986), (2) Carilli et al. (1998), (3) Snellen et al.~(2000).}
\end{center}
\end{table}
For each of these we performed a spectral scan along the
line-of-sight. The high sensitivity and large bandwidth (1 GHz),
combined with the possibility of observing simultaneously with two
receivers, permitted us to scan a range of $\approx10$ GHz in both the
2-mm and 3-mm bands. Over these ranges we reached optical depth limits
in both bands sensitive enough the detect the 4 known redshifted
millimetre absorbers (Table 1 cf. $\tau\approx0.7{\rm ~to~}\approx2$
at $\gapp4$ \kms ~resolution), although no $\geq3\sigma$ absorption
features were found (Murphy, Curran \& Webb 2003).

Of the remaining visually dim quasars, 10 have 3-mm flux densities
$\gapp0.5$ Jy. Four of these are located in the north and with the
NRO\footnote{The fact that each of the 6 AOSs on the NRO only covers
0.25 GHz is compensated by the high efficiency of the 45-m antenna (4
Jy K$^{-1}$ cf. 25 Jy K$^{-1}$ at SEST).} we were able to observe the
three listed in Table 1.  While the 6-mm limits are poor, again at
3-mm our search is sensitive enough to detect the 4 known absorbers
over the observed redshift range: For the $J=0\rightarrow1$,
$1\rightarrow2$ and $2\rightarrow3$ of transitions CO, HCN and
HCO$^+$, i.e. the most commonly detected transitions in the 4 known
absorbers, the observed frequencies give a 50\% coverage for 0742+103
up to $z\approx3$ and 30\% for both 1600+335 and 1655+077 up to the
emission redshift\footnote{Note that we have included the possibility
of HCN or HCO$^+$ $0\rightarrow1$ Galactic absorption towards all 3
sources as well as HCN or HCO$^+$ $1\rightarrow2$ in the host of
1600+335.}. The coverage for the SEST sources are discussed in Murphy,
Curran \& Webb (2003); for the above transitions up to 90\% is achieved
due to the large bandwidth and dual receiver capability.









\end{document}